\documentclass[]{emulateapj}

\begin{document}

\title{Spectrally resolved pure rotational lines of water in protoplanetary disks}

\author{Klaus M. Pontoppidan\altaffilmark{1}}
\author{Colette Salyk\altaffilmark{1,2}}
\author{Geoffrey A. Blake\altaffilmark{1}}
\author{Hans Ulrich K{\"a}ufl\altaffilmark{3}}

\altaffiltext{1}{California Institute of Technology, Division of Geological and Planetary Sciences, 
MS 150-21, Pasadena, CA 91125; pontoppi@gps.caltech.edu}
\altaffiltext{2}{The University of Texas at Austin, Department of Astronomy, 1 University Station C1400, Austin, Texas 78712, USA}
\altaffiltext{3}{European Southern Observatory, Karl-Schwarzschild-Strasse 2, 85748 Garching bei München, Germany}

\begin{abstract}
We present ground-based high resolution N-band spectra ($\Delta v = 15\,\rm km\,s^{-1}$) of pure rotational lines of water vapor in
two protoplanetary disks surrounding the pre-main sequence stars AS 205N and RNO 90, selected based on detections
of rotational water lines by the Spitzer IRS. Using VISIR on the Very Large Telescope, we spectrally resolve individual lines and show that they
have widths of 30-60\,$\rm km\,s^{-1}$, consistent with an origin in Keplerian disks at radii of $\sim 1\,$AU. The water lines have similar widths to those of the CO at 4.67 $\mu$m, indicating that the mid-infrared water lines trace similar radii.   The rotational 
temperatures of the water are $540$ and $600$\,K in the two disks, respectively. However, the lines ratios show evidence of non-LTE excitation, with low-excitation line fluxes being
over-predicted by 2-dimensional disk LTE models. Due to the limited number of observed lines and the non-LTE line ratios, an accurate measure of the water ortho/para ratio is not available, but a best estimate for AS 205N is ortho/para $=4.5\pm1.0$, apparently ruling out a low-temperature origin of the water. 
The spectra demonstrate that high resolution spectroscopy of rotational water lines is feasible from the ground, and further that ground-based high resolution spectroscopy is likely to significantly improve our understanding of the inner disk chemistry recently revealed by recent Spitzer observations. 
\end{abstract}

\keywords{astrochemistry --- planetary systems: protoplanetary disks}

\section{Introduction}

The study of the dynamics and chemistry of the inner ($R<10\,$AU) regions of disks around young pre-main sequence stars
is a field in rapid growth, to a large extent thanks to the availability of sensitive infrared instrumentation, including the InfraRed Spectrograph (IRS)
on the Spitzer Space Telescope. The molecular gas at these radii plays a key role in the process of planet formation, not only as a necessary reservoir for the formation of
gas giants, but also in the generation and eventual delivery of volatile molecular species to terrestrial planets. 

Recent observations have demonstrated that the mid-infrared wavelength range in protoplanetary disks contains a forest of molecular emission lines, many
of them due to pure rotational transitions of water vapor \citep{Carr08,Salyk08,Pascucci09,Pontoppidan10}. 
While \cite{Carr04} detected and spectrally resolved rovibrational water lines at $2.3\,\mu$m and \cite{Salyk08} at 3\,$\mu$m from protoplanetary disks, 
the observations of the mid-infrared pure rotational lines were all carried out at $R=\lambda/\Delta\lambda\le 600$, resulting in unresolved
and blended line spectra. The rovibrational 2-3\,$\mu$m lines have much higher excitation temperatures, and may not probe the
same gas as the mid-infrared rotational lines. Further, they also suffer from significant line blending, making it difficult to study the 
details of the excitation structure. In this Letter, we demonstrate that high resolution spectroscopy of {\it pure rotational transitions of } water in the atmospheric $N$-band window
is possible. The rotational water lines observable from the ground have excitation temperatures of 3000-6000\,K, directly
comparable to those of the well-studied CO rovibrational lines in the 4.7\,$\mu$m fundamental band \citep{Najita03,Blake04,Brittain07,Pontoppidan08,Salyk09}, 
allowing for more accurate relative abundance measurements. Further, the lines are well separated from one another and do not suffer from blending with other species. 

In this Letter, we present high resolution ($R\sim 20\,000$) $N$-band spectra of AS 205N and RNO 90, disks known to have strong water vapor emission
 \citep{Salyk08, Pontoppidan10}, obtained using
the Very Large Telescope Imager and Spectrometer for the mid-InfraRed \citep[VISIR,][]{Lagage04}. Both disks are likely part of the
Ophiuchus star forming cloud at a distance of 120\,pc \citep{Loinard08}. With stellar luminosities of 4.0 and 3.5\,$L_{\odot}$ \citep{Chen95} and spectral
types of K5 and G5, the central stars have likely masses close to 1\,$M_{\odot}$ \citep[e.g., ][]{Siess97, Salyk08, Pontoppidan10b}. RNO 90 in particular, being a single star, may be a fairly 
close analog to the young Sun, while AS 205N is the primary of a 1\farcs3 binary (160\,AU).

\section{Observations}

Six water lines with excitation energies varying from
from 3300-5800\,K were selected using the list of \cite{Pontoppidan09} in combination with
the limitations imposed by the availability of filters on VISIR. While telluric water generates highly pressure-broadened 
absorption, we find that even from a relatively low site, such as Paranal Observatory (2635\,m), the atmospheric transmission for all our targeted lines is $>$30\%, and often significantly better.

Spectra centered on 12.405 and 12.454\,$\mu$m were observed using the cross-dispersed mode of VISIR, with a chop-throw of 8\arcsec. 
In this mode, one source position is off the slit because the minimum chop-throw is larger than the slit length. Additional settings
centered on 12.829 and 12.893\,$\mu$m were observed using the long-slit mode, in which all source positions are on slit, doubling the effective
exposure time. The effective on source exposure times were 1250\,s and 2500\,s for the cross-dispersed and long-slit modes for
AS 205N, respectively. For RNO 90, these exposure times were doubled. All observations were carried out using the 0\farcs75 slit during good seeing conditions, 
specifically ensuring that components of the AS 205 binary are well separated. AS 205N was observed on August 5 and 6, 2009, while RNO 90 was observed on separate nights 
between August 9, 2009 and September 11, 2009. 
The spectral images were co-added using the ESO VISIR pipeline version 3.2.2. Chopping parallel to the slit removes most of the sky background, but a significant
residual, especially from rapidly varying atmospheric water lines, is still visible. These residuals were removed by subtracting the median of each row (in the cross-dispersion
direction) from the spectral image. The 1D spectra were generated using optimal extraction \citep{Horne86}. The wavelength calibration from the ESO pipeline was used, and 
is computed by cross correlating the sky emission spectrum with an atmospheric model generated using HITRAN \citep{Rothman05}.

\begin{figure}
\centering
\includegraphics[width=8cm]{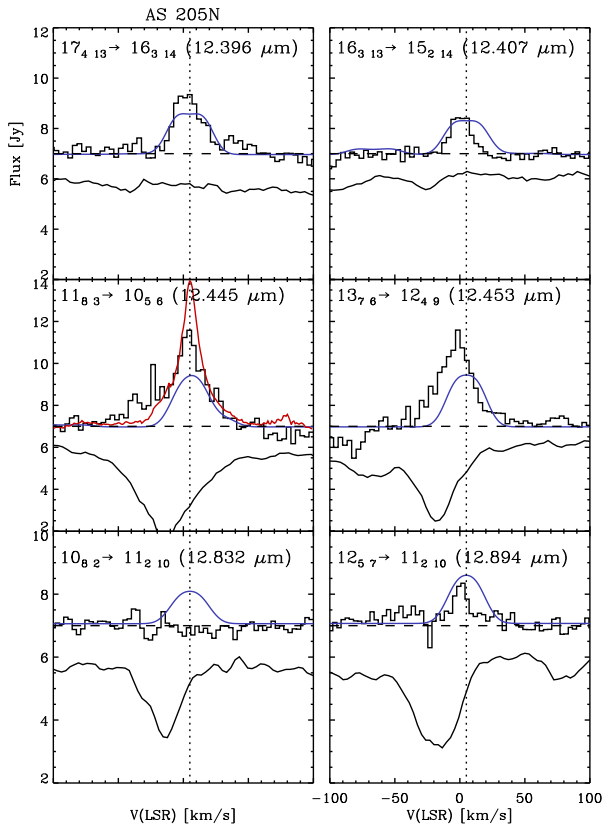}
\caption[]{VISIR spectra of rotational water lines observed in AS 205N. The curves below the spectra
show the standard star spectra, which are a combination of the spectral response function and the atmospheric transmission. The velocity
range of the spectra is referenced to the local standard of rest. The red curve is the CO rovibrational ($\Delta v=1$) line shape ($J<8$), as
observed with CRIRES \citep{Pontoppidan10b}, whose center defines the vertical dotted line. The blue curves are lines calculated using an LTE 2D disk model (see text for details).
}
\label{AS205}
\end{figure}

\begin{figure}
\centering
\includegraphics[width=8cm]{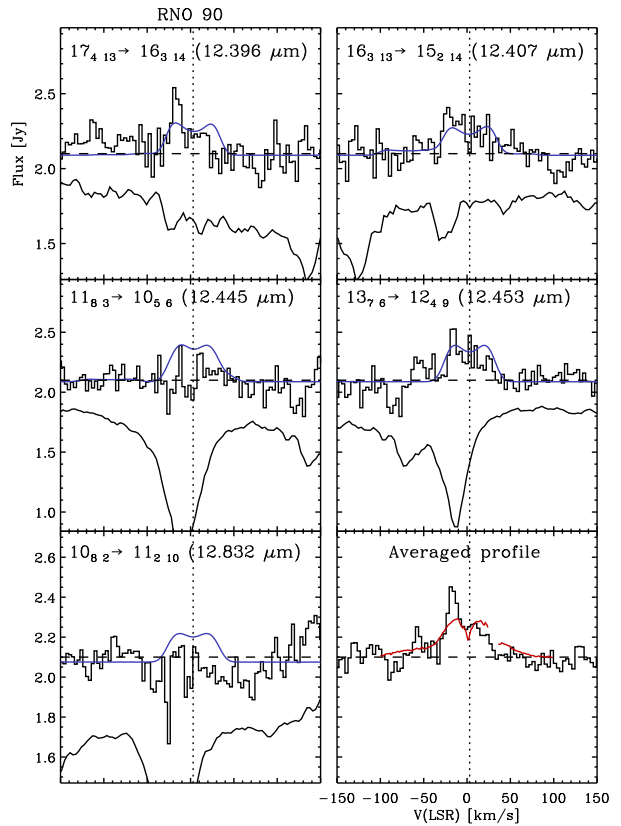}
\caption[]{Same as Figure \ref{AS205}, but for RNO 90. The lower right panel presents the average of the three detected lines. The asymmetry visible as an increase in flux on the
blue side of the detected lines is possibly real, but since it coincides with the telluric water lines, we are hesitant to discuss it in detail without further observations, preferably with different
reflex motion doppler shifts. }
\label{RNO90}
\end{figure}

The spectral response function and atmospheric transmission spectrum were measured using observations of a bright infrared excess source known to 
not display strong water emission at 12\,$\mu$m in the survey of \cite{Pontoppidan10}, specifically the bright Herbig Ae/Be star HD150193. Ideally, 
a hot star without excess should be used, but no such source is sufficiently bright in the N-band. One could use a later type star, but in that case photospheric absorption
lines will be present that may be difficult to remove. Hence, the choice was made to use an excess source dominated by continuum emission and with no
apparent lines. Corrections for small shifts (of order a few pixels) between the source spectrum and that of the standard as well 
as a correction for differences in water column densities using a simple Lambert-Beer law were made to minimize the 
telluric residuals. 
The absolute fluxes were scaled to match those of the Spitzer spectra in \cite{Pontoppidan10}.

For comparison with the water lines, we use archival M-band high resolution (R=100,000) spectra of the CO rovibrational fundamental ($\Delta v=1$) lines observed with CRIRES. The
CO lines have roughly the same upper level energies as the N-band rotational water lines. The CRIRES
data set will be described in greater detail in a spectro-astrometric survey \citep{Pontoppidan10b}. It was
processed following the procedures of \cite{Pontoppidan08}. Finally, we extracted an isolated lower energy line for comparison from the \cite{Pontoppidan10} Spitzer spectra. 
The low-resolution (R=600) Spitzer spectra do not allow for the measurement of many unblended lines, but there are a few exceptions, 
one of them being the $10_{4~7}\rightarrow 9_{3~6}$ ortho line at 33.5\,$\mu$m. 

\section{Results}

The reduced VISIR spectra of AS 205N and RNO 90 are shown in Figures \ref{AS205} and \ref{RNO90}, respectively. 
For AS 205N, we detect 5 out of the 6 targeted lines, while 3 lines are detected for RNO 90. Integrated line fluxes are given in
Table \ref{obs_table}.

The lines, where detected, are spectrally resolved, with line widths of $\sim$$30\,\rm km\,s^{-1}$ for AS 205N and $\sim$$60\,\rm km\,s^{-1}$ 
for RNO 90. In the figures, the water lines are compared to the CO $\Delta v$=1 line shape. While
variations are apparent for AS 205N, these differences are relatively minor, with FWHM of 20-35\,$\rm km\,s^{-1}$. similar to, or
slightly less than those of CO. This indicates that the water lines are formed at similar or slightly larger disk radii.  
The high energy lines ($17_{4~3}\rightarrow 16_{3~4}$, $16_{3~13}\rightarrow 15_{2~14}$) of AS 205N show indications of
being flat-topped, while the lower excitation lines are single peaked, as is CO. 
All three detected RNO 90 lines are consistent with being double peaked, also matching the corresponding CO lines. The
double peak is apparent when the three detected line profiles are averaged. We interpret this as evidence for a line origin in a Keplerian disk.

\begin{deluxetable}{lllll}
\tablecaption{H$_2$O line fluxes}
\tablehead{
\colhead{Line}   & \colhead{Wavel.} & \colhead{$E_{\rm upper}$} & \colhead{AS 205N\tablenotemark{a}} & \colhead{RNO 90\tablenotemark{a}} \\
                       &  \colhead{$\mu$m} & \colhead{K} & &
}
\startdata
$17_{4~13}\rightarrow 16_{3~14}$ (o)& 12.396 & 5746 &$4.8\pm 0.2$  &$0.7\pm 0.08$\\
$16_{3~13}\rightarrow 15_{2~14}$ (p) & 12.407 & 4915 &$2.3\pm 0.2$  &$0.9\pm 0.11$\\
$13_{7~6}\rightarrow 12_{4~9}$ (o) & 12.453 & 4187 &$10.4\pm 0.2$ &$1.0\pm 0.14$\\
$11_{8~3}\rightarrow 10_{5~6}$ (o) & 12.445 & 3606 &$11.4\pm 0.3$  &$<0.3$\\
$12_{5~7}\rightarrow 11_{2~10}$ (p) & 12.893 & 3290 &$1.9\pm 0.2$    &$<0.3$\\
$10_{8~2}\rightarrow 9_{5~5}$ (p) & 12.832 & 3223 &$< 0.4$             &$<0.2$\\
$10_{4~7}\rightarrow 9_{3~6}$ (o)  & 33.501 & 2260 &$1.9\pm 0.2$    &$<0.3$
\enddata
\tablenotetext{a}{$10^{-14}\,\rm erg\,cm^{-2}\,s^{-1}$}
\label{obs_table}
\end{deluxetable}

\begin{figure}
\includegraphics[width=8cm]{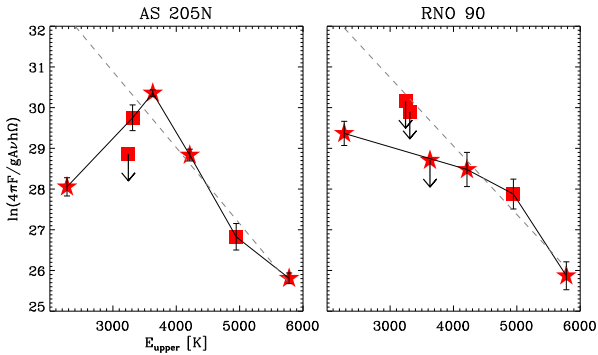}
\caption[]{Rotation diagram for AS 205N and RNO 90. The star symbols are ortho lines, while the square symbols indicate para lines. The points include
a spin degeneracy of 3:1. The dashed lines are
the best fit single temperature models (using the detected VISIR lines only). The units are cgs and the symbols have their usual meaning. }
\label{rotdia}
\end{figure}

\begin{figure}
\includegraphics[width=8cm]{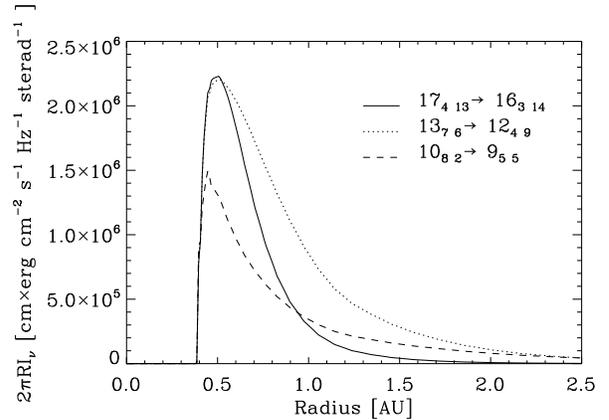}
\caption[]{Continuum-subtracted surface brightness profiles of three lines for the RNO 90 model. }
\label{emit_radius}
\end{figure}

In order to estimate the properties of the emitting gas (rotational temperatures, radial distribution, column densities and abundances), we model the spectra using the 2D raytracer RADLite \citep{Pontoppidan09}.
We also construct rotation diagrams (Figure \ref{rotdia}), which are designed to allow for easy comparison with a model consisting of a single-temperature slab of gas, and enable a discussion of the excitation and column density, in an average sense, of the observed species; LTE level populations and optically thin emission predicts that line fluxes arrange along
a straight line in the rotation diagram. Here we demonstrate that there are departures from the basic slab model assumptions of LTE, and that the assumption
that all lines are formed in the same solid angle does not hold for water. Specifically, the slab model predicts 
significantly brighter $10_{8~2}\rightarrow 9_{5~5}$ and $12_{5~7}\rightarrow 11_{2~10}$ emission, by at least factors of 5-10, than what is observed for AS 205N. The upper limits on the 
RNO 90 lines are not strongly constraining, and the rotation diagram for this disk is roughly consistent with the slab model in the N-band, while the
Spitzer 33.5\,$\mu$m line is underluminous. The lower energy Spitzer line thus confirms the N-band turnover at lower energies in the rotation diagram for AS 205N. 
We stress that these are differences that are not apparent from the low-resolution Spitzer spectra alone. 
A linear regression of the VISIR lines on the linear part of the rotation diagram ($E_{\rm upper}>3500\,$K) yields rotational temperatures of 540 and 600\,K for AS 205N and RNO 90, 
respectively, for optically thin emission. 

The AS 205N temperature can be compared to the 1000\,K required to fit the 3\,$\mu$m rovibrational lines \citep{Salyk08}, 
indicating that the rotational lines indeed trace cooler gas. The selection of lines includes ortho- and para-transitions. The assumption that
the N-band lines should form a straight line in the rotation diagram allows an estimate to be made of the ortho/para ratio. The paucity of lines, 
and the presence of clear departures from LTE and/or slab geometry unfortunately makes such an estimate uncertain. Nevertheless, scaling the
three ortho-lines for AS 205N with energies between 3500 and 6000\,K to an optimal alignment with the two detected para lines results in an apparent O/P ratio of $4.5\pm1.0$ for AS 205N The ratio is higher than 3 because the para lines are placed below a straight line defined by the three ortho lines, but by less than $2\sigma$.  
While this measurement is certainly affected by the caveats discussed above, we interpret this as indicative of a warm gas-phase (equilibrium) chemistry, as opposed to water 
formed in a low-temperature environment ($<$60\,K) via grain-surface chemistry, which would produce O/P ratios $<$3. Indeed, cometary water
has O/P ratios corresponding to spin temperatures of $\sim$30\,K \citep{Crovisier97, Kawakita04}, and optical depth effects would drive the
estimated O/P values lower.

Since the lines are now spectrally resolved, the emitting area and hence the column density of the emitting
gas can be directly estimated by converting the line width to a physical location in a Keplerian disk. The inclination of the RNO 90 disk is $\sim$$45\degr$ \citep{Pontoppidan10b} and the star has a mass of
$\sim$$0.9\,M_{\odot}$ \citep{Andrews09}. The RNO 90 water lines extend to $\sim 30\,\rm km\,s^{-1}$, corresponding to
an inner water line emitting radius of 0.45\,AU. Similarly, for AS 205N the line width at zero flux corresponds to an inner emitting 
radius of 0.1-0.4\,AU, for the $11_{8~3}\rightarrow 10_{5~6}$ and $17_{4~13}\rightarrow 16_{3~14}$ lines, respectively, assuming Keplerian flow, a disk 
inclination of $25\degr$ and a stellar mass of 1\,$M_{\odot}$ \citep{Andrews09}. The effective outer radii of the emitting areas are determined by 
generating 2D models for the lines with RADLite, assuming LTE level populations. The model uses a simple parametrized flared disk structure 
\citep{Dullemond01}, with inner radii
fixed according to the observed line widths. The water abundance is set to $2.6\times 10^{-4}$ per H and the disk surface gas-to-dust ratio set to $1.28\times 10^{4}$, simulating significant
dust settling, as discussed in \cite{Meijerink09}. Finally, the stellar luminosities were adjusted to match the dust continuum at 12.5\,$\mu$m. The resulting line strengths and profiles
match well to those of the detected N-band lines with $E >$4000\,K, as was already suggested by the rotation diagram (see Figures \ref{AS205} and \ref{RNO90}).
Figure \ref{emit_radius} shows the corresponding line surface brightness from the RADLite model, demonstrating that the RNO 90 emission is dominated by radii from 0.4 to 1.0\,AU
($\sim$$2.6\,\rm AU^2$), but
with the emitting area changing by a factor 2-3, depending on the line. Due to the higher luminosity of AS 205N, emitting radii span a wider range, here from 0.35 to 2.5\,AU 
($\sim$$19\,\rm AU^2$). Using these
measures, the column densities of water, averaged over the emitting radii, are $2\times 10^{18}$ and $3\times 10^{18}\,\rm cm^{-2}$. 

\section{Discussion}

We have detected pure rotational lines from water vapor (H$_2$$^{16}$O) in protoplanetary disks around T Tauri stars from the ground using transitions in the atmospheric $N$-band window. 
Gas temperatures of 500-600\,K are found, while the line widths of 30-60\,$\rm km\,s^{-1}$ confirm that the disk surface at radii of 0.4-2.5\,AU is forming the lines. 
The resolved lines enable a much more direct determination of the emitting area and hence the column density of the water, as compared to that possible for
spectrally unresolved Spitzer observations. We find that the assumptions used for the analysis of the Spitzer data generally hold, i.e. that the lines are formed in the surface of Keplerian disks at
radii of $\sim$1\,AU. The water column densities are found to be a little higher for AS 205N than those determined by \cite{Salyk08} using rotational lines at somewhat longer wavelengths and lower 
energies. This appears to be consistent with the rotation diagram that indicates that lower excitation lines are less bright than expected for LTE level populations. This effect was also noted by \cite{Meijerink09}, 
who interpreted it as a depletion of water beyond $\sim$1\,AU. More extensive N-band studies will help to further constrain the shape of the water rotation diagram to constrain the excitation
and spatial distribution of water vapor in the inner regions of protoplanetary disks. 

The derived water abundance is close to that predicted by \cite{Bethell09}, based on a chemical model that includes water self-shielding. 
The 2D RADLite models of both disks require high water abundances and high gas-to-dust ratios in order to reproduce the observed lines -- a significant fraction of the available oxygen must be locked up
in water vapor, and substantial dust settling to the disk midplane appears to have already taken place. An initial estimate of the ortho/para ratio of the water suggests a value consistent with
a high temperature equilibrium (O/P=3), but higher signal-to-noise observations of more lines is required to confirm a high O/P ratio.
If confirmed, this is indicative of water formed via hot gas-phase chemistry, as opposed to water initially formed as ice in the outer disk and transported inwards as part of migrating solids. 

The spectra demonstrate that it is possible to obtain high resolution spectra of individual rotational lines of water in protoplanetary disks 
from the ground, in spite of telluric water absorption. These new tracers of the warm molecular layer have several unique advantages: The excitation energies
and collisional rates of the high rotational levels are very similar to those of the ubiquitous fundamental ro-vibrational band of CO at 4.7\,$\mu$m. Given the chemical
stability of CO, this allows for detailed studies of the relative water abundances in a well-defined location in the disk surface. The lines are strong, also relative to the continuum, and their
analysis is not hampered by strong photospheric lines from the central star -- in contrast to the weaker, much higher energy water lines in the 3\,$\mu$m ``hot band'' \citep{Salyk08}. 
The $N$-band lines may provide our only opportunity to obtain detailed velocity information of water at $\sim 1-10\,$AU in disks around young solar analogs for the foreseeable future. 
Their observability from the ground also make them available to the very high spatial resolution line imaging of 10m class telescopes, as well as the next generation of Extremely Large Telescopes. 
As such, they will be important complements to the lower resolution spectroscopy of the James Webb Space Telescope. 

\acknowledgments{Based on observations made at the ESO Paranal Observatory 
under program IDs 084.C-0635 and 179.C-0151. The authors acknowledge valuable discussions with Alain Smette. 
}

\bibliographystyle{apj}

\end{document}